\documentclass[11pt]{article}
\usepackage{setspace}
\usepackage{amsmath}
\usepackage{graphicx}
\usepackage{epstopdf}
\usepackage[cm]{fullpage}
\usepackage{subfig}
\usepackage{blkarray}
\usepackage{caption}
\usepackage[numbers,compress,sort]{natbib}
\usepackage[]{nomencl}
\usepackage{authblk}
\usepackage{braket}
\DeclareMathOperator{\Tr}{Tr}

\title{Phonon Eigenspectrum-Based Formulation of the Atomistic Green's Function Method}
\author[1]{Sridhar Sadasivam\footnote{present address: Center for Nanoscale Materials, Argonne National Laboratory, Argonne, IL 60439}}
\author[2]{Umesh V. Waghmare}
\author[1]{Timothy S. Fisher\footnote{present address: Mechanical and Aerospace Engineering Department, University of California, Los Angeles, Los Angeles, CA 90095 email: tsfisher@ucla.edu}}
\affil[1]{School of Mechanical Engineering and Birck Nanotechnology Center, Purdue University, West Lafayette, IN 47907}
\affil[2]{Theoretical Sciences Unit, Jawaharlal Nehru Centre for Advanced Scientific Research, Jakkur, Bangalore 560064}
\date{\vspace{-10ex}}
\begin{document}
\maketitle
\section*{Abstract}
While the atomistic Green's function (AGF) method has the potential to compute spectrally resolved phonon transport across interfaces, most prior formulations of the AGF method provide only the total phonon transmission function that includes contributions from all phonon branches or channels. In this work, we present a formulation of the conventional AGF technique in terms of phonon eigenspectra that provides a natural decomposition of the total transmission function into contributions from various phonon modes. The method involves the use of Dyson and Lippmann-Schwinger equations to determine surface Green's functions from the phonon eigenspectrum of the bulk, and establishes a direct connection between the transmission function and the bulk phonon spectra of the materials forming the interface. We elucidate our formulation of the AGF technique through its application to a microscopic picture of phonon mode conversion at Si-Ge interfaces with atomic intermixing. Intermixing of atoms near the interface is shown to increase the phase space available for phonon mode conversion and to enhance thermal interface conductance at moderate levels of atomic mixing. The eigenspectrum-based AGF (EAGF) method should be useful in determination of microscopic mechanisms of phonon scattering and identification of the specific modes that dominate thermal transport across an interface.
 \section{Introduction}
The ability to resolve details of heat conduction in solids to the level of individual phonon polarizations, or branches, is becoming more important to the understanding and engineering of thermal transport, particularly with the emergence of commensurate experimental techniques that target coherent phonon behavior. Polarization-specific transport has been postulated to dictate important phenomena such as the effects of boundary roughness in dimensionally confined nanomaterials \cite{feser2012thermal}, transport across interfaces \cite{singh2011effect}, and thermoelectric devices \cite{nolas1999skutterudites}, among others. At the same time, phonon transport modeling beyond the historically common Debye approximation has become much more accessible to researchers through the use of first-principles calculations of atomic structure and dynamics \cite{baroni2001phonons}. Here, we present an extension of the atomistic Green's function method that separates contributions from individual phonon branches in a computationally efficient manner, with application to interfacial thermal transport. 

Resolution of phonon contributions to thermal conductivity from frequency (energy) and mean-free path has become more common in recent research \cite{qiu2012molecular,regner2013broadband}. For example, the so-called `thermal conductivity accumulation' function provides insights into how domain boundaries can affect the measured conductivity \cite{yang2013mean}. Similarly, frequency resolution is particularly important for understanding phonon transport across interfaces \cite{huang2010simulation}. For both frequency- and scattering-resolved transport, each non-degenerate phonon polarization can produce markedly different behavior. Consequently, any such studies should ideally distinguish among the phonon branches. Ultimately, high levels of spectral and scattering resolution can be employed in general device-scale solvers, such as those based on the Boltzmann transport equation with complex scattering processes \cite{minnich2011quasiballistic}. 

The atomistic Green's function (AGF) method has proven to be very effective in simulating ballistic phonon transport in crystalline materials, beginning with the early work by Mingo and Yang \cite{mingo2003phonon}. The approach is particularly well suited to the study of phonon transport across interfaces, such as nanoscale contacts \cite{mingo2009green} and dimensionally mismatched interfaces \cite{huang2010simulation}. The method has also been extended to include anharmonic scattering, albeit with substantial computational complexity \cite{mingo2006anharmonic}. However, the basic AGF methodology groups all the active phonon branches together in the calculation of the frequency-dependent transmission function, which forms the basis of thermal transport calculations within the Landauer framework. Some prior studies have reported methods to distinguish contributions from individual phonon branches within the transmission process \cite{huang2011modeling,ong2015efficient}, and a comparison of these approaches with that presented in this paper is provided in Section \ref{comparison_prior}.  

The main virtue of the AGF method is associated with the high spectral fidelity of its transport predictions, particularly when compared to alternatives such as molecular dynamics with wave packets \cite{schelling2002phonon}. Here, we report a computationally efficient approach to achieve polarization-specific AGF predictions of spectral phonon transport. The subsequent sections describe the essential elements of the method, the  treatment of surface Green's functions using Dyson and Lippmann Schwinger equations, and the extraction of polarization-specific transmission functions. The paper ends with an example of transport across a heterojunction interface to demonstrate the efficacy of the method and general conclusions. 

\section{Atomistic Green's Function Method: A Brief Review}
This section provides a brief overview of the AGF method, and the reader is referred to prior literature \cite{zhang2007atomistic,sadasivam2014atomistic} for a more detailed introduction to the method. The AGF technique is ideally suited to perform calculations of transport across a `device' region connected to semi-infinite `contacts'. The central quantity of interest is the device Green's function $G_d$ given by:
\begin{equation}
G_d(\omega;q_{t}) = [\omega^2I-H_d-\Sigma_1-\Sigma_2]^{-1}
\end{equation}
where $H_d$ denotes the force constant matrix corresponding to the device region, and $\Sigma_1$, $\Sigma_2$ represent the self-energies due to the semi-infinite contacts. $q_{t}$ denotes the in-plane or transverse wavevector (perpendicular to the transport direction). The self-energies are obtained from the surface Green's functions $g_1$, $g_2$ of the contacts as follows:
\begin{equation}
\Sigma_1(\omega;q_{t}) = \tau_1g_1\tau_1^{\dagger} \qquad \Sigma_2(\omega;q_{t}) = \tau_2g_2\tau_2^{\dagger}
\end{equation}
where $\tau_1$, $\tau_2$ represent the interaction between device and contacts, i.e., force constant matrices for bonding between atoms in the device region with atoms in the left and right contacts, respectively. The procedure for obtaining the surface Green's functions $g_1$, $g_2$ is discussed in the next section. Additional matrices $\Gamma_1$, $\Gamma_2$ are obtained from the imaginary part of the self-energy matrices $\Sigma_1$, $\Sigma_2$.
\begin{equation}
\Gamma_1(\omega;q_{t}) = i(\Sigma_1-\Sigma_1^\dagger) \qquad \Gamma_2(\omega;q_{t}) = i(\Sigma_2-\Sigma_2^\dagger)
\end{equation}
$\Gamma_1$, $\Gamma_2$ are termed `escape rates' \cite{sadasivam2014atomistic}, because they physically represent the rate of transfer of phonons between the contacts and the device. The transmission function $\mathcal{T}(\omega)$ from contact 1 to contact 2 is obtained from the Caroli formula:
\begin{equation}
\mathcal{T}(\omega;q_{t}) = \Tr[\Gamma_1G_d\Gamma_2G_d^{\dagger}]
\label{Caroli_formula} 
\end{equation}
The total thermal conductance $G_Q$ across the device region is obtained by integration over all phonon frequencies and transverse wavevectors \cite{zhang2007simulation}:
\begin{equation}
G_{Q} = \sum\limits_{q_{t}}{\frac{1}{2\pi}\int\limits_{0}^{\infty}{\hbar\omega}\mathcal{T}(\omega,q_{t})\frac{\partial f_{BE}^o}{\partial T}d\omega}
\end{equation}
\section{Computation of Surface Green's Functions}
\label{surf_g}
The surface Green's functions $g_1$, $g_2$ are typically obtained using the Sancho-Rubio technique that is also referred to as the  decimation method in prior literature \cite{guinea1983effective,sancho1985highly}. The Sancho-Rubio method is an iterative scheme that provides quick convergence of the surface Green's function because the number of contact layers considered in the calculation increases exponentially with the number of iterations. However, the decimation technique provides only the total surface Green's function and is not suited for calculating contributions of individual phonon eigenmodes to the surface Green's function. Hence, this method cannot be used to obtain mode-resolved transmission functions and the contribution of individual phonon modes to the interface thermal conductance. In this section, we present an alternative technique for the computation of surface Green's functions using the bulk phonon eigenmodes of the contacts. The present technique leads naturally to the definition of polarization-resolved surface Green's functions that can be used in the AGF method to calculate polarization-resolved transmission functions. 
\subsection{Calculation of Bulk Green's Function}
\begin{figure}
\begin{center}
\includegraphics[width=160mm]{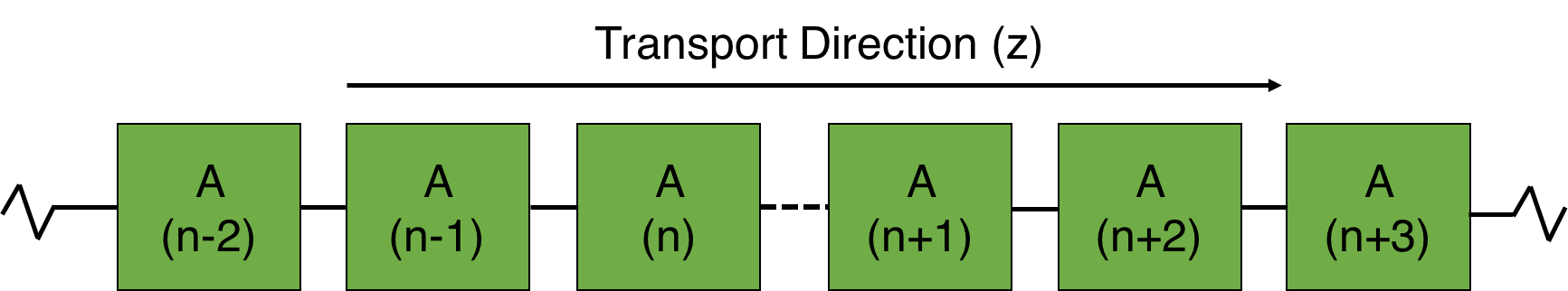}
\caption{Schematic of a bulk periodic material. Two semi-infinite contacts are created by removing the connection between the $n^{th}$ and the $(n+1)^{th}$ unit cells.}
\label{bulk_A}
\end{center}
\end{figure}
Consider a bulk material represented by a repeating unit `A' as shown in Figure \ref{bulk_A}. The repeating unit `A' could in general contain multiple atoms and could also be periodic in the transverse direction. For a given phonon wavevector $\boldsymbol{q}$ ($\boldsymbol{q} = \boldsymbol{q}_{t}+q_z \hat{z}$), the corresponding frequencies and eigenvectors are obtained by diagonalization of the phonon dynamical matrix $D(\boldsymbol{q})$:
\begin{equation}
\omega^2\phi = (H_{on}(\boldsymbol{q_{t}})+\tau_l(\boldsymbol{q_{t}})\exp{(-iq_za)}+\tau_r(\boldsymbol{q_{t}})\exp{(iq_za)})\phi = D(\boldsymbol{q})\phi
\label{normal_eig}
\end{equation}
where $H_{on}$ denotes the on-site element of the dynamical matrix and $\tau_l$, $\tau_r$ denote the connections to the left and right nearest neighbors respectively. The term $a$ denotes the periodicity along transport direction and is chosen large enough to ensure that only nearest neighbor interactions exist. The foregoing equation corresponds to the usual method of obtaining phonon dispersion and requires the solution of a normal eigenvalue problem. 

Now we consider the inverse problem of obtaining the phonon wavevector (in the transport direction) and the eigenvector for a given phonon frequency $\omega$ and transverse wavevector $q_{t}$. This problem requires the solution of a generalized eigenvalue problem twice the size of the normal eigenvalue problem in Eq.~(\ref{normal_eig}) \cite{khomyakov2005conductance,luisier2006atomistic}: 
\begin{equation}
\left({\begin{array}{cc}
\omega^2I-H_{on}(\boldsymbol{q_{t}}) & -\tau_r(\boldsymbol{q_{t}}) \\ I & 0\\ \end{array}}\right)\left({\begin{array}{c} \phi_i \\ \phi_{i+1}\\ \end{array}} \right) = 
e^{-iq_za}\left({\begin{array}{cc}
\tau_l(\boldsymbol{q_{t}}) & 0 \\ 0 & I\\ \end{array}}\right)\left({\begin{array}{c} \phi_i \\ \phi_{i+1}\\ \end{array}} \right)
\label{gevp}
\end{equation}  
The solution of Eq.~(\ref{gevp}) can also generate complex wavevectors, in addition to real wavevectors that are typically used in a phonon dispersion calculation. The real wavevectors correspond to propagating Bloch waves, while the complex wavevectors physically represent evanescent waves that decay to the left or right depending on the sign of the imaginary part of the wavevector. Such complex wavevectors always occur in conjugate pairs, i.e., if $q_z$ is a complex wavevector that satisfies the above equation, $q_z^*$ is also a solution of the above equation. After solving for the wavevectors and eigenvectors that satisfy the generalized eigenvalue problem, the bulk Green's function is obtained from the following expression:
\begin{equation}
\begin{split}
G_{m,n}(\omega;\boldsymbol{q_{t}}) & = -ia\left(\sum\limits_{q_z^+}\frac{\phi(\boldsymbol{q_{t}},q_z^+)\phi^{\dagger}(\boldsymbol{q_{t}},q_z^{+*})}{2\omega v_{g,z}(\boldsymbol{q_{t}},q_z^+)}\exp{[i(m-n)q_za]}\Theta(m-n)\right) \\  & +ia\left(\sum\limits_{q_z^-}\frac{\phi(\boldsymbol{q_{t}},q_z^-)\phi^{\dagger}(\boldsymbol{q_{t}},q_z^{-*})}{2\omega v_{g,z}(\boldsymbol{q_{t}},q_z^{-})}\exp{[i(m-n)q_za]}\Theta(n-m)\right)
\end{split}
\label{bulk_G}
\end{equation}
The derivation of Eq.~(\ref{bulk_G}) involves the use of contour integration in the complex wavevector plane and is an extension to phonons of the Green's function expression for electrons derived in refs.~\cite{allen1979green,chang1982complex}. The first (second) summation on the right side runs over all real wavevectors with a positive (negative) group velocity $v_{g,z}$ along the z-direction and all imaginary wavevectors with a positive (negative) imaginary part. $\phi$ denotes the normalized phonon eigenvector corresponding to one repeating unit. $\Theta(m)$ represents the Heaviside step function ($\Theta(0)$ is taken to be 0.5) and is used to ensure that Bloch waves with positive (negative) group velocity and evanescent waves with positive (negative) imaginary parts contribute to the left (right) off-diagonal element of the Green's function. The indices $m$, $n$ denote the block indices of the Green's function, i.e., $G_{m,m}$ corresponds to an on-site element of the bulk Green's function matrix while $G_{m,m-1}$, $G_{m,m+1}$ correspond to off-diagonal elements of the Green's function matrix. Translational symmetry along the z-direction dictates that the matrices $G_{m,m}$, $G_{m,m-1}$, $G_{m,m+1}$ are independent of the value of $m$. The group velocity $v_{g,z}(\boldsymbol{q_{t}},q_z)$ can be obtained from Hellmann-Feynman theorem as \cite{mcgaughey2014predicting}:
\begin{equation}
v_{g,z}(\boldsymbol{q_{t}}q_z) = \frac{\phi^\dagger(\boldsymbol{q_{t}},q_z^*)\left(\partial D(\boldsymbol{q_{t}},q_z)/\partial q_z\right)\phi(\boldsymbol{q_{t}},q_z)}{2\omega_{\boldsymbol{q}}}
\end{equation}
where $\partial D(\boldsymbol{q_{t}}q_z)/\partial q_z = -ia\tau_l(\boldsymbol{q_{t}})\exp{(-iq_za)}+ia\tau_r(\boldsymbol{q_{t}})\exp{(iq_za)}$ denotes the derivative of the dynamical matrix with respect to the wavevector along the transport direction. 
\subsection{Surface Green's Function from Dyson's Equation}
If $G^o$ corresponds to the Green's function for a Hamiltonian $H^o$, Dyson's equation \cite{mingo2009green} can be used to obtain the Green's function $G$ of a modified Hamiltonian $H$ where $H = H^o+V$ ($V$ denotes the perturbation to the original Hamiltonian $H^o$):
\begin{equation}
G = G^o+G^oVG
\end{equation}
The foregoing equation is applied to obtain the surface Green's function $g$ from the bulk Green's function $G$ computed in the previous section. The relevant perturbation to the force constant matrix essentially involves setting elements of the bonds between two adjacent unit cells to zero (as shown by the dotted line between the $n^{th}$ and $(n+1)^{th}$ repeating units in Figure \ref{bulk_A}). The only non-zero terms in the perturbation matrix $V$ correspond to $V_{n,n+1}=-\tau_r$ and $V_{n+1,n}=-\tau_l$. Application of Dyson's equation leads to the following expressions for the surface Green's functions of the left and right contacts:
\begin{equation}
g_{n,n} = (I+G_{n,n+1}\tau_l)^{-1}G_{n,n} \qquad g_{n+1,n+1} = (I+G_{n+1,n}\tau_r)^{-1}G_{n+1,n+1}
\label{dyson_g}
\end{equation}
When the left and right contacts constitute different materials, the corresponding bulk Green's functions and harmonic matrices of the left and right contacts need to be used in the above equations, i.e., the left (right) surface Green's function $g_{n,n}$ ($g_{n+1,n+1}$) will depend on the bulk Green's functions and harmonic matrices of the left (right) contact. The present procedure constitutes an alternative method to obtain surface Green's functions in which the phonon eigenvectors of the bulk contacts are explicitly used in the construction of bulk Green's functions, and Dyson's equation is invoked to obtain the surface Green's function from the bulk Green's function.
\section{Polarization-Resolved Surface Green's Functions}
\label{pol_g}
The algorithm presented in the previous section involving the use of Dyson's equation is useful to obtain the total surface Green's function. However, the computation of mode-resolved transmission functions requires the calculation of each bulk phonon mode's contribution. In this section, we present an algorithm involving the use of the Lippmann-Schwinger equation to obtain the surface Green's functions directly from the bulk eigenvectors of the contacts. Analogous to the Dyson's equation, the Lippmann-Schwinger equation can be used to obtain the eigenvectors $\psi$ corresponding to a perturbed Hamiltonian $H=H^o+V$ from the Green's function $G$ of the perturbed Hamiltonian and eigenvectors $\phi$ of the original Hamiltonian \cite{mingo2009green}:
\begin{equation}
\psi = \phi+GV\phi
\end{equation}
where $\phi$ denotes the eigenvectors of the original Hamiltonian that corresponds to the bulk contact in our model, and $\psi$ denotes the eigenvectors of the modified Hamiltonian that corresponds to two semi-infinite surfaces. Similar to the previous section, the only non-zero elements of the perturbation matrix $V$ are the connections between the $n^{th}$ and $(n+1)^{th}$ unit cells. Application of the Lippmann-Schwinger equation results in the following expressions for the surface eigenvectors $\psi_{n}$ and $\psi_{n+1}$ of the left and right semi-infinite contacts respectively:
\begin{equation}
\psi_n = \phi_n-g_{n,n}\tau_r\phi_{n+1} \qquad \psi_{n+1} = \phi_{n+1}-g_{n+1,n+1}\tau_l\phi_{n}
\end{equation}
where $g_{n,n}$ and $g_{n+1,n+1}$ correspond to the total surface Green's functions of the left and right contacts respectively. Only eigenvectors with real wavevectors, i.e., propagating Bloch waves, contribute to heat conduction across the interface. Hence, only the surface eigenvectors corresponding to real wavevectors of the original unperturbed Hamiltonian are computed. Because the eigenvectors of the unperturbed Hamiltonian corresponding to real wavevectors are Bloch waves, $\phi_{n+1} = \phi_n\exp{(iq_za)}$ in the above equation. The surface eigenvector corresponding to a propagating Bloch wave of the bulk contact can be used to compute the contribution of that phonon mode to the surface Green's function. For instance, if $\psi_{n,\alpha}$ and $\psi_{n+1,\beta}$ are the surface eigenvectors corresponding to propagating eigenvectors $\phi_\alpha$ and $\phi_\beta$ of the left and right contacts respectively, then the surface Green's functions $g_{1,\alpha}$ and $g_{2,\beta}$ corresponding to these modes can be written as:
\begin{equation}
g_{1,\alpha}(\omega;\boldsymbol{q_{t}}) = -\frac{ia}{2}\frac{\psi_{n,\alpha}\psi_{n,\alpha}^\dagger}{2\omega v_{g,z}(\boldsymbol{q_{t}}q_{z,\alpha})} \qquad g_{2,\beta}(\omega;\boldsymbol{q_{t}}) = \frac{ia}{2}\frac{\psi_{n+1,\beta}\psi_{n+1,\beta}^\dagger}{2\omega v_{g,z}(\boldsymbol{q_{t}}q_{z,\beta})}
\label{surf_g_pol}
\end{equation}
The foregoing equation assumes that the mode $\phi_\alpha$ of the left contact propagates in the positive z-direction (towards the device, $v_{g,z}(\boldsymbol{q_{t}},q_{z,\alpha})>0$), and the mode $\phi_\beta$ of the right contact contact propagates in the negative z-direction (towards the device region, $v_{g,z}(\boldsymbol{q_{t}},q_{z,\beta})<0$). Eq.~(\ref{surf_g_pol}) is analogous to Eq.~(\ref{bulk_G}) used to construct the bulk Green's function; the factor of 2 in the denominator of the above equations derives from the Heaviside step function in Eq.~(\ref{bulk_G}) where $\Theta(0)=1/2$. Such polarization-resolved surface Green's functions can be used to define polarization-resolved self-energies $\Sigma_{1,\alpha} = \tau_1g_{1,\alpha}\tau_1^\dagger$, $\Sigma_{2,\beta} = \tau_2g_{2,\beta}\tau_2^\dagger$ and escape rate matrices $\Gamma_{1,\alpha} = i(\Sigma_{1,\alpha}-\Sigma_{1,\alpha}^\dagger)$, $\Gamma_{2,\beta} = i(\Sigma_{2,\beta}-\Sigma_{2,\beta}^\dagger)$. The transmission function from phonon mode $\phi_{\alpha}$ of the left contact to the phonon mode $\phi_{\beta}$ of the right contact is given by the Caroli formula of Eq.~(\ref{Caroli_formula}), but with mode-specific escape rate matrices $\Gamma_{1,\alpha}$ and $\Gamma_{2,\beta}$. 
\begin{equation}
\mathcal{T}_{\alpha\beta}(\omega,\boldsymbol{q_{t}}) = \Tr[\Gamma_{1,\alpha}G_d\Gamma_{2,\beta}G_d^\dagger]
\end{equation}
\section{Comparison with Prior Methods}
\label{comparison_prior}
\begin{figure}
\begin{center}
\includegraphics[height=80mm]{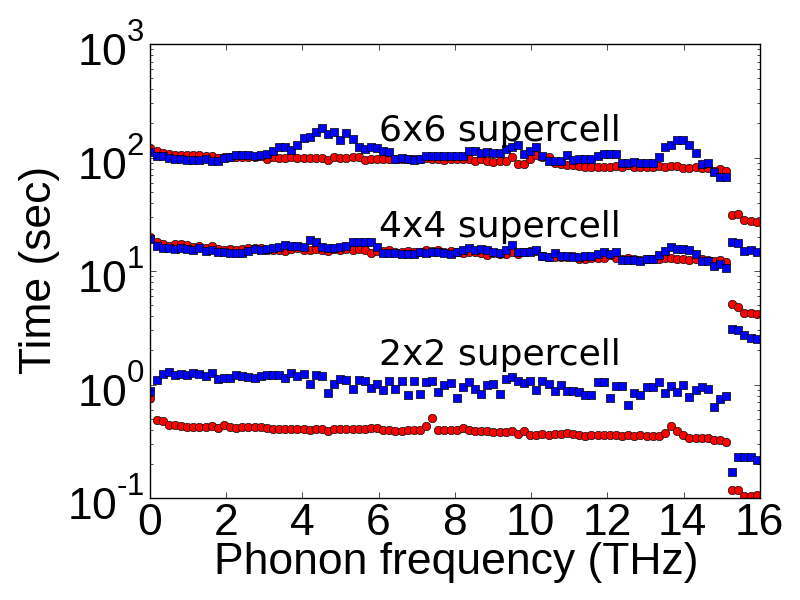}
\caption{Computational time for obtaining mode-resolved transmission coefficients using the Bloch matrices approach of ref.~\cite{ong2015efficient} (red circles) and the algorithm presented in Section \ref{pol_g} of this paper (blue squares). The computational times are shown as a function of phonon frequency and supercell size.}
\label{comp_performance}
\end{center}
\end{figure}

Huang et al.~\cite{huang2011modeling} developed a method to compute polarization-resolved transmission functions in the AGF framework using eigendecomposition of the imaginary part of the surface Green's function matrix ($a_{1,2} = i(g_{1,2}-g_{1,2}^{\dagger})$). However, the surface Green's function must still be computed using the Sancho-Rubio method, and the eigenvectors of the matrix $a_{1,2}$ do not have an explicit relation with the bulk phonon eigenmodes of the contacts. Ong and Zhang \cite{ong2015efficient} proposed a mode-matching approach to compute polarization-resolved transmission functions through a direct connection between bulk phonon eigenvectors and the transmission function. While the method of ref.~\cite{ong2015efficient} provides the same physical quantities as the formulation proposed here, the eigenspectrum approach provides a few advantages as highlighted below:
\begin{itemize}
\item The eigenspectrum method has been shown to be significantly (60-100 times faster) more efficient \cite{luisier2006atomistic} than the decimation/Sancho-Rubio method for electron transport. The efficiency gain is obtained through transformation of the generalized eigenvalue problem of Eq.~(\ref{gevp}) into a normal eigenvalue problem. This transformation has not been implemented for the computations in this paper and we only solve the generalized eigenvalue problem. However, such techniques from electron transport \cite{luisier2006atomistic} point towards the potential for increased computational efficiency of the eigenspectrum approach over the widely used decimation technique.  
\item The only difference between the present method and the conventional AGF technique implemented in prior literature is the direct use of bulk phonon modes to compute surface Green's functions instead of the more commonly employed decimation technique. The eigenspectrum approach provides a natural decomposition of contact self-energies into contributions from eigenmodes and illustrates that the conventional AGF technique for total transmission provides all necessary ingredients for also obtaining mode-resolved transmission coefficients with the only difference of an alternate technique to evaluate surface Green's functions. In contrast to the method proposed in ref.~\cite{ong2015efficient}, the present formulation is only a slight modification of the widely employed conventional AGF method with the same mathematical formulation for mode-resolved and total transmission functions. 
\item In Figure \ref{comp_performance}, we compare the computational time for obtaining mode-resolved transmission functions from Si to Ge as a function of phonon frequency (transverse wavevector at $\Gamma$) for different supercell sizes (see next section for details of the simulation setup). To compare the computational efficiency of the section of the algorithm for mode-resolved transmission functions between the technique in this work and that in ref.~\cite{ong2015efficient}, the decimation technique is used to obtain the surface Green's function in both approaches, i.e., the total surface Green's function was not obtained using the approach in Section \ref{surf_g} rather from the decimation technique. The total surface Green's function is then used in the formulation of Section \ref{pol_g} to obtain mode-resolved transmission coefficients. This procedure effectively provides a comparison of the computational performance of the Bloch matrices approach and the mode-resolved Green's function approach without any differences in the technique for obtaining the total surface Green's function. We find that both approaches require similar computational times for all frequencies and simulation domain sizes indicating that the Bloch matrices approach and the mode-resolved surface Green's functions approach produce similar computational scaling. The mode-resolved surface Green's function technique however shows slight variation of the computational time with phonon frequency depending on the number of propagating eigenmodes at each frequency. 
\end{itemize}
\section{Demonstration}
In this section, the foregoing technique to compute mode-resolved transmission functions is applied to phonon transport across Si-Ge interfaces with varying levels of intermixing between Si and Ge atoms at the interface. This problem has previously been studied \cite{li2012effect,tian2012enhancing} using the conventional AGF method, but we demonstrate that the present mode-resolved approach provides further insights into the effects of atomic intermixing on interfacial phonon scattering. Similarly to ref.~\cite{tian2012enhancing}, we consider a simplified description of the Si-Ge interface in which the lattice constants of both Si and Ge lattices are set to the bulk lattice constant of Si, i.e., the effect of lattice mismatch on phonon transmission is not considered. Also, the present AGF formulation is applicable to elastic phonon transport, and anharmonic scattering processes are not modeled in this approach. Prior molecular dynamics simulations have shown that anharmonic effects do not significantly affect the thermal interface conductance of Si-Ge interfaces for temperatures less than 500 K \cite{landry2009thermal}. 

\begin{figure}
\centering
\subfloat[]{\includegraphics[height=40mm]{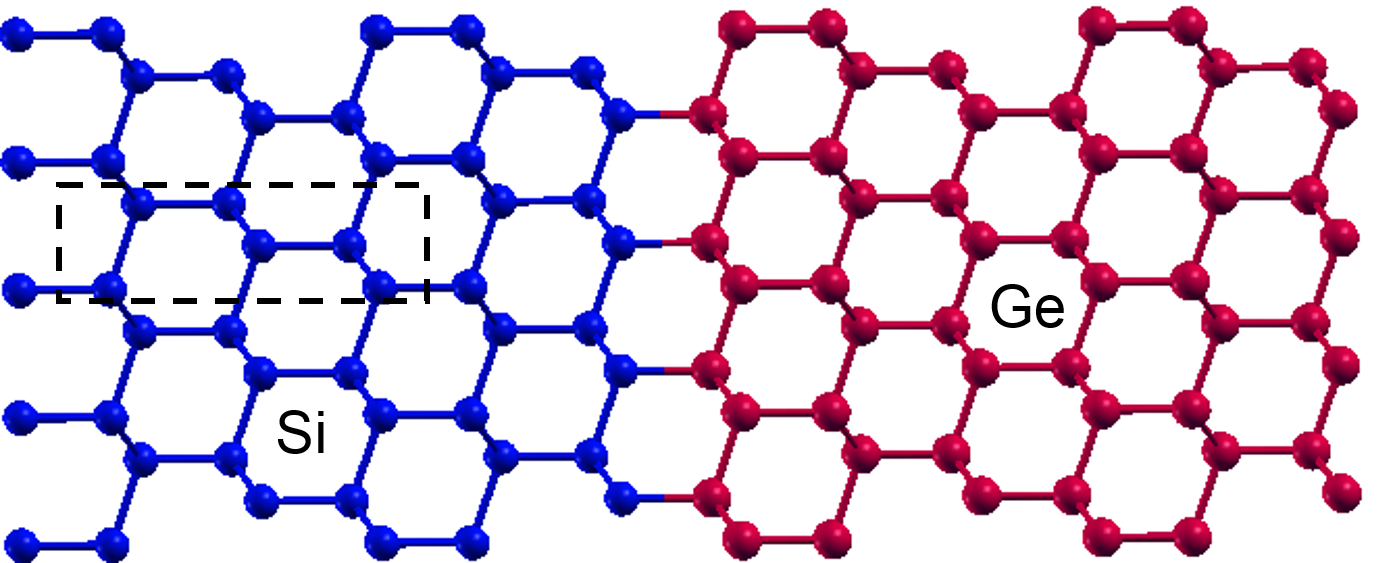}}\\
\subfloat[]{\includegraphics[height=60mm]{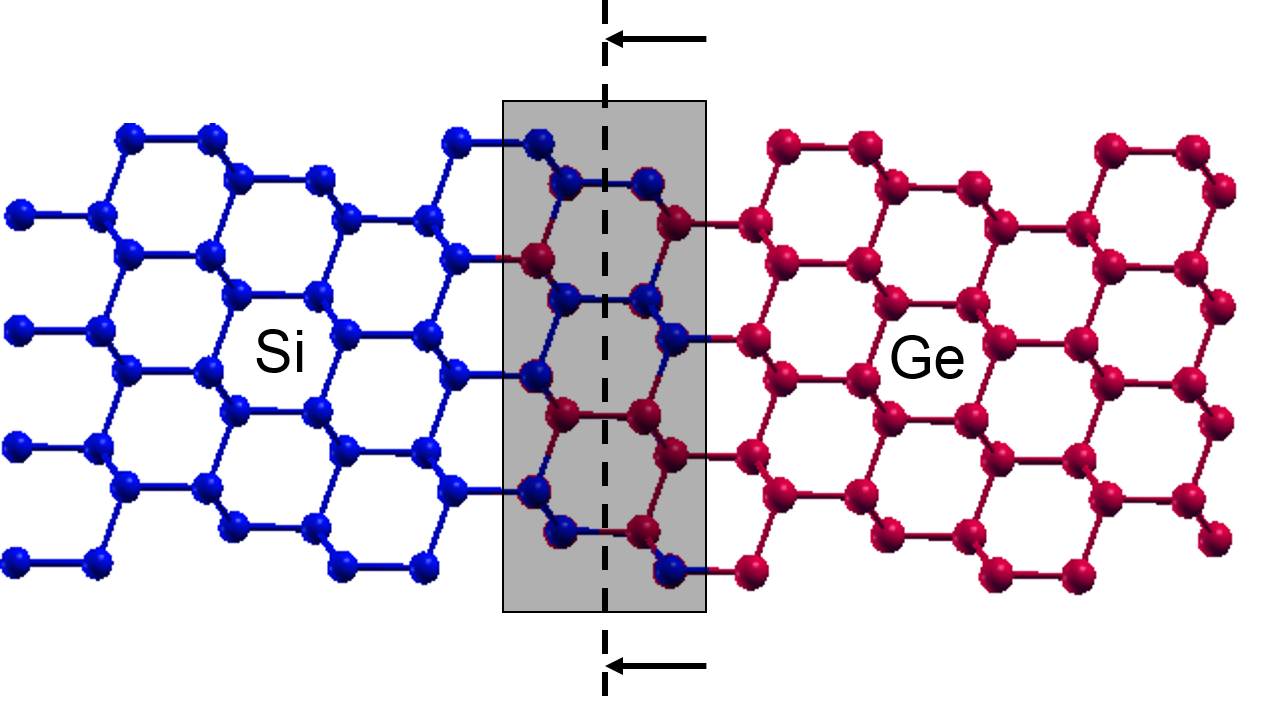}}\qquad\quad
\subfloat[]{\includegraphics[height=60mm]{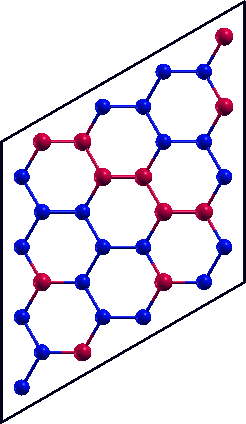}}
\caption{a) Ideal Si-Ge interface with no intermixing of atoms. The dotted box corresponds to a six-atom non-primitive unit cell used in DFPT calculations of phonon dispersion. b) Si-Ge interface with four layers of random atomic intermixing (shaded rectangular box) c) Cross-sectional view of the intermixed interface (the dotted line in (b) indicates the cross-sectional plane used for visualization).}\label{Si_Ge_schematic}
\end{figure}

Figure \ref{Si_Ge_schematic}a shows the atomic structure of an ideal Si-Ge interface oriented along the [111] direction. Apart from the ideal or smooth interface shown in Figure \ref{Si_Ge_schematic}a, Si-Ge interfaces with varying levels of random atomic intermixing are also considered in the phonon transport simulations. The protocol for generating atomic structures with intermixing of Si and Ge atoms is similar to the approach adopted in ref.~\cite{tian2012enhancing}. We define an intermixing length around the interface, and all atoms within this region are randomly assigned the mass of Si or Ge with a uniform probability. As an example, Figures \ref{Si_Ge_schematic}b and \ref{Si_Ge_schematic}c show an intermixed Si-Ge interface with an intermixing length of four atomic layers (shaded rectangular box around the interface). All results reported in this paper are averaged over three random realizations of intermixed atomic structures. 
\begin{figure}
\begin{center}
\includegraphics[height=60mm]{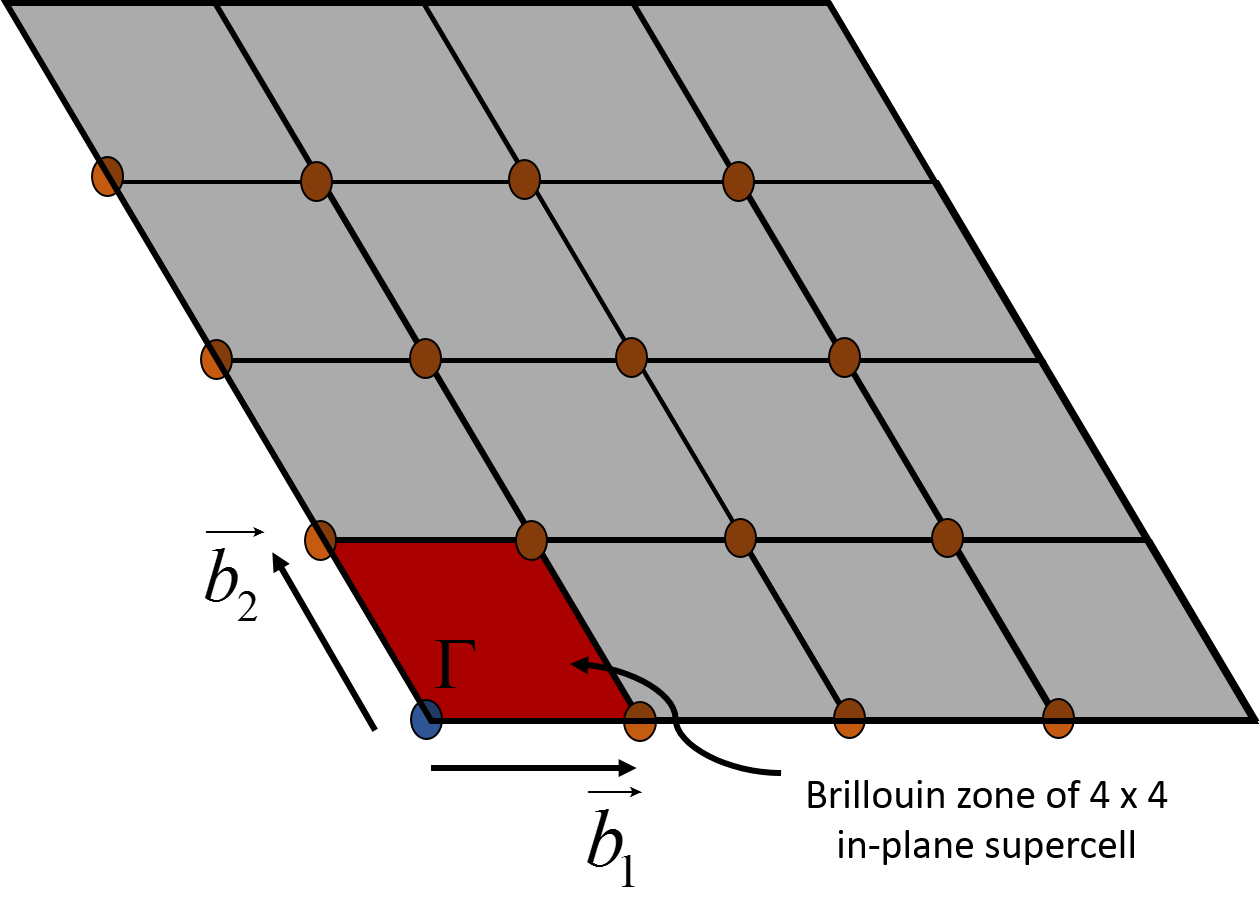}
\caption{The large grey shaded parallelogram represents the BZ of a primitive unit cell in the transverse plane. The small red shaded parallelogram represents the BZ of a $4\times4$ supercell. The $\Gamma$ point in the BZ of the supercell (blue circle) contains zone-folded modes from all the q-points points represented by orange circles in the BZ of the primitive unit cell.}
\label{inplane_BZ}
\end{center}
\end{figure}

The interatomic force constants of Si are obtained from density functional perturbation theory (DFPT) using Rappe-Rabe-Kaxiras-Joannopoulos (RRKJ) ultrasoft pseudopotentials, and the exchange correlation energy is approximated under the generalized gradient approximation (GGA) using the Perdew-Burke-Ernzerhof (PBE) functional form. A six-atom non-primitive unit cell (see black dotted box in Figure \ref{Si_Ge_schematic}a) oriented along the [111] direction is considered in the DFPT calculations with a planewave kinetic energy cutoff of 680 eV and an electron k-point grid of $12\times 12\times 9$ in self-consistent calculations of charge density. The real-space interatomic force constants are computed from a Fourier transform of the dynamical matrices obtained on a $4\times 4\times 3$ q-point grid used for the linear response calculations. The lattice vectors $\vec{a}_1$, $\vec{a}_2$, $\vec{a}_3$ of the unit cell shown in Figure\ref{Si_Ge_schematic}a are given by:
\begin{equation}
\vec{a}_1 = \frac{a}{\sqrt{2}} \hat{y} \qquad \vec{a}_2 = \frac{\sqrt{3}a}{2\sqrt{2}}\hat{x}+\frac{a}{2\sqrt{2}}\hat{y} \qquad \vec{a}_3 = \sqrt{3}a\hat{z}
\end{equation}
where the relaxed lattice constant obtained from DFT is $a = 5.46$ $\text{\AA}$. Similar to ref.~\cite{tian2012enhancing}, the force constants of Ge are assumed to be the same as that of Si, effectively implying that the mass mismatch between Si and Ge is the only factor affecting the phonon transmission function. This simplifying assumption is used to demonstrate the new approach developed in this paper and to obtain a qualitative understanding of the effects of atomic intermixing on interfacial phonon mode conversion. Quantitatively accurate calculations must consider the effect of changes in the local force field at the interface due to atomic inter-mixing \cite{gu2015phonon}.

All simulations in this paper consider a $4 \times 4$ in-plane supercell within which atoms are randomly intermixed, i.e., the in-plane lattice constants of the supercell are given by $4\vec{a}_1$, $4\vec{a}_2$. The in-plane supercell corresponds to 192 atoms (16 in-plane  unit cells of 12 atoms each) within each block shown in Figure \ref{bulk_A}. Hence, all the matrices in Eq.~(\ref{dyson_g}) are of size $576\times 576$. Due to zone folding, the modes at the $\Gamma$ point of the $4 \times 4$ supercell Brillouin zone include modes from multiple q-points (see orange circles in Figure \ref{inplane_BZ}) of the BZ of the primitive cell (extended zone scheme). The transmission function obtained from the conventional AGF approach provides only the total transmission probability summed over all these different zone-folded modes. However, the present extension to mode-resolved transmission functions enables a rigorous analysis of individual transmission coefficients. 

\begin{figure}
\begin{center}
\includegraphics[height=60mm]{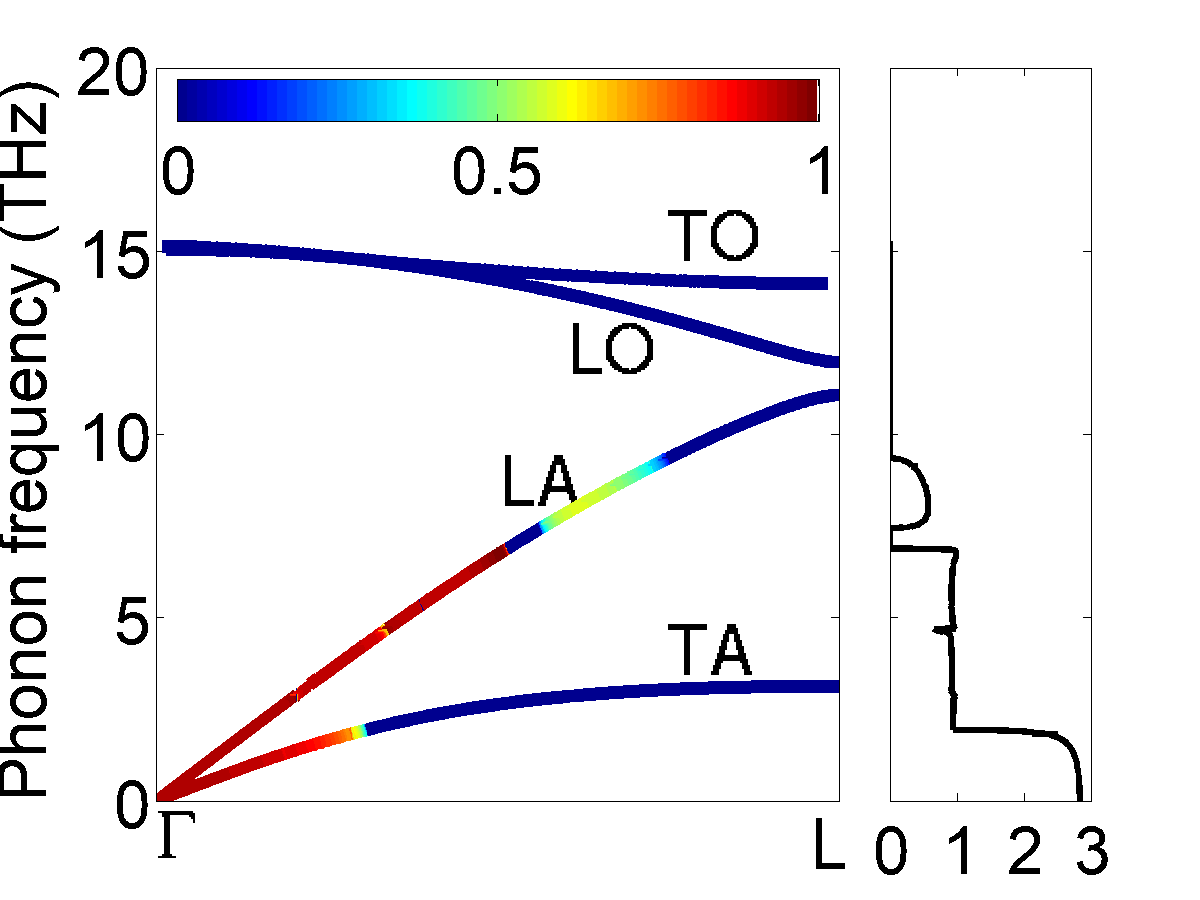}
\caption{The left panel shows the phonon dispersion of Si along the [111] direction with individual phonon modes colored according to transmission probability across a smooth Si-Ge interface. The right panel shows the total transmission function of normally incident phonon modes as a function of phonon frequency. }
\label{mode_resolved_trans}
\end{center}
\end{figure}

As an example, Figure \ref{mode_resolved_trans} shows the transmission probabilities of normally incident phonon modes in Si across a smooth Si-Ge interface, i.e., the transverse wavevector is the $\Gamma$ point of the primitive unit cell. Both TA and LA modes of Si exhibit near-unity transmission until the TA (2.5 THz) and LA (7 THz) cutoff frequencies of Ge along the [111] direction. The LA mode also exhibits mode conversion into the optical modes of Ge with a reduced transmission probability of approximately 0.5 in the frequency range 8 to 10 THz. All the Si phonon modes above the maximum phonon frequency in Ge exhibit zero transmission.  

We also compared our predictions of transmission probability at normal incidence with the Bloch matrices technique of ref.~\cite{ong2015efficient}, and found that the Bloch matrices approach does not predict equal transmission coefficients for the degenerate TA modes (the predictions for LA mode transmission matched correctly between the two methods). We found that diagonalization of Bloch matrices to calculate eigenvectors does not produce the correct orthogonal pair of eigenvectors for degenerate phonon modes. The present technique, however, predicts equal transmission for degenerate phonon modes as expected from symmetry considerations.

\begin{figure}
\centering
\subfloat[]{\includegraphics[height=70mm]{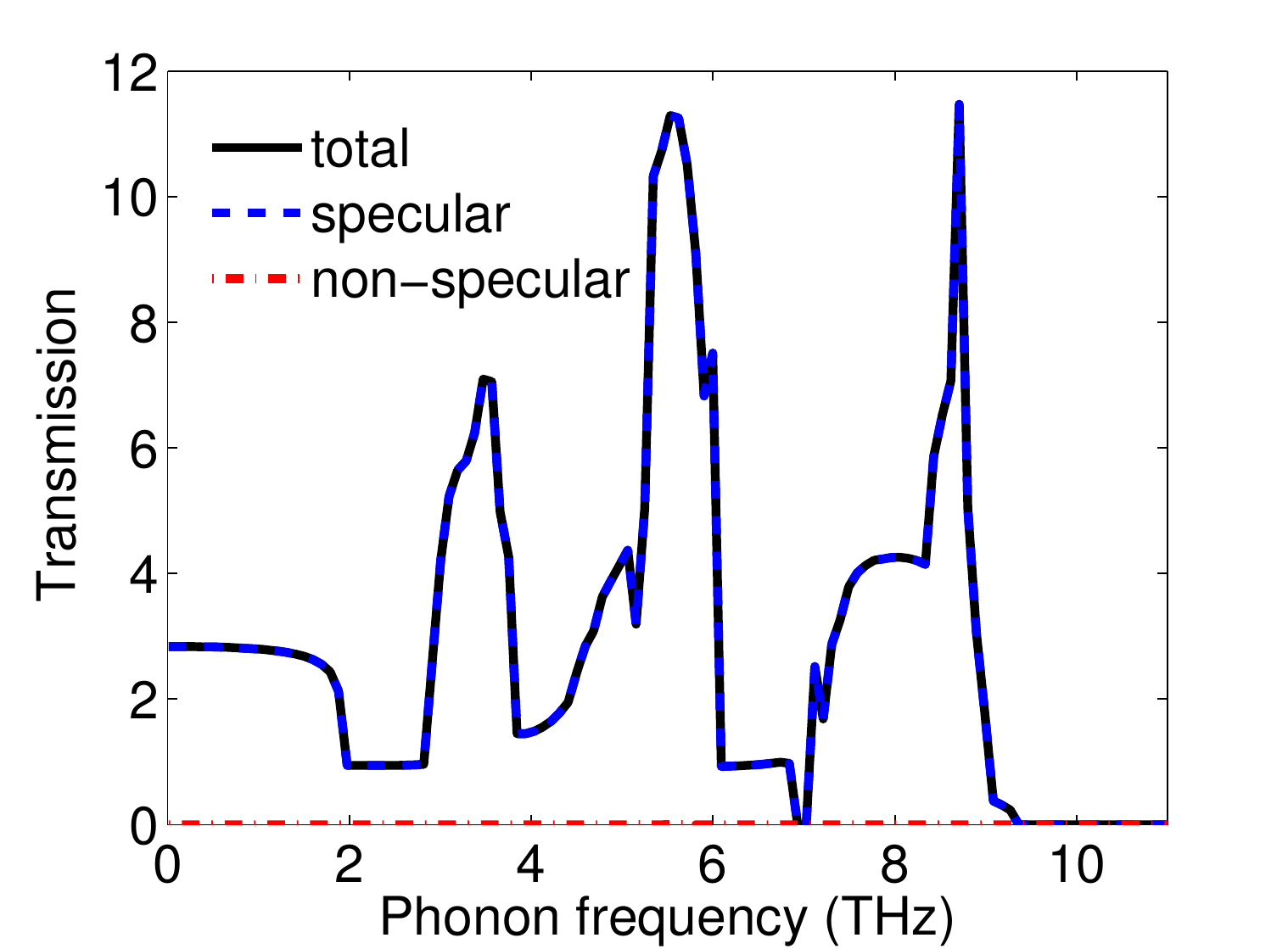}}
\subfloat[]{\includegraphics[height=70mm]{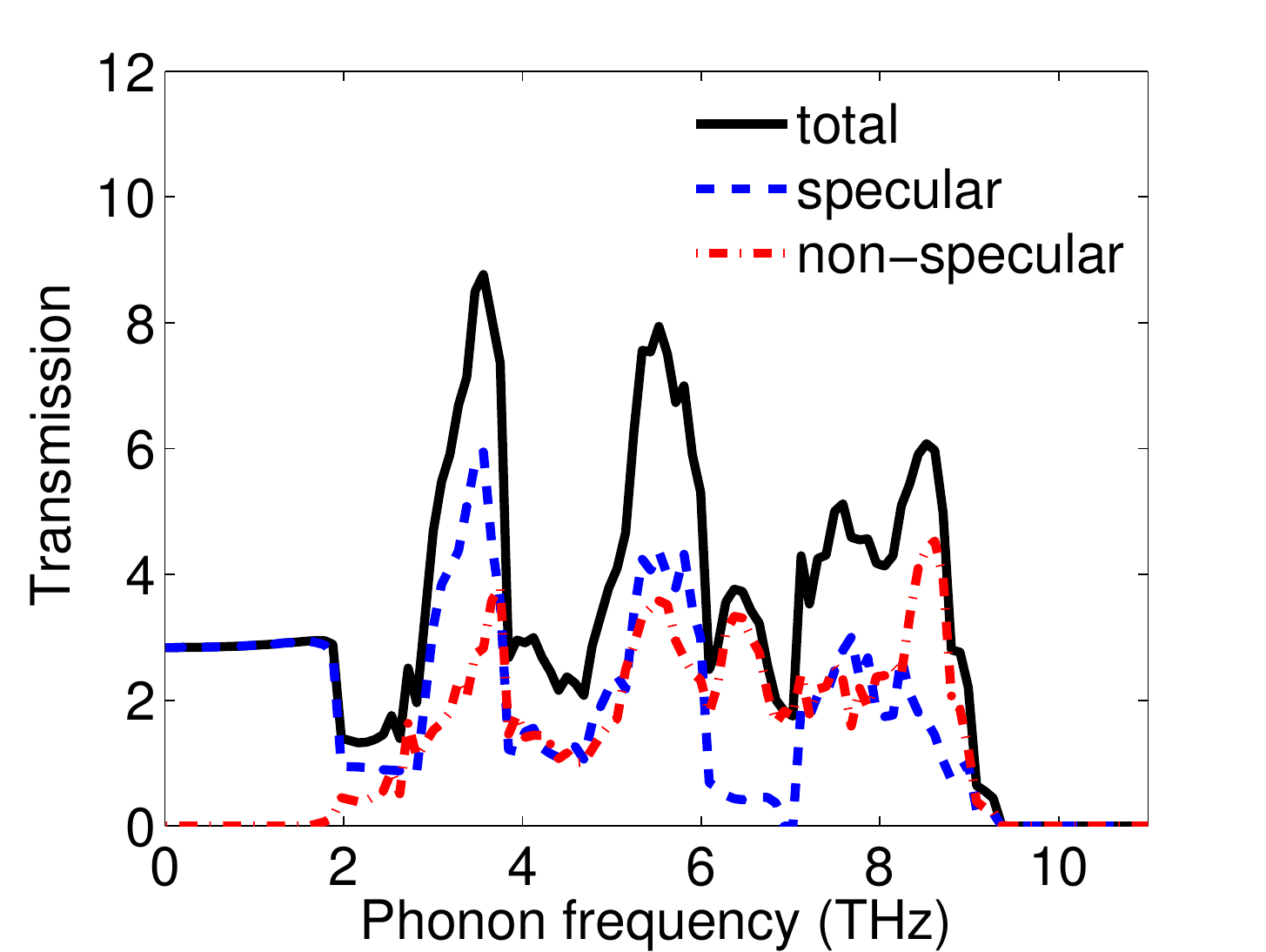}}\\
\subfloat[]{\includegraphics[height=70mm]{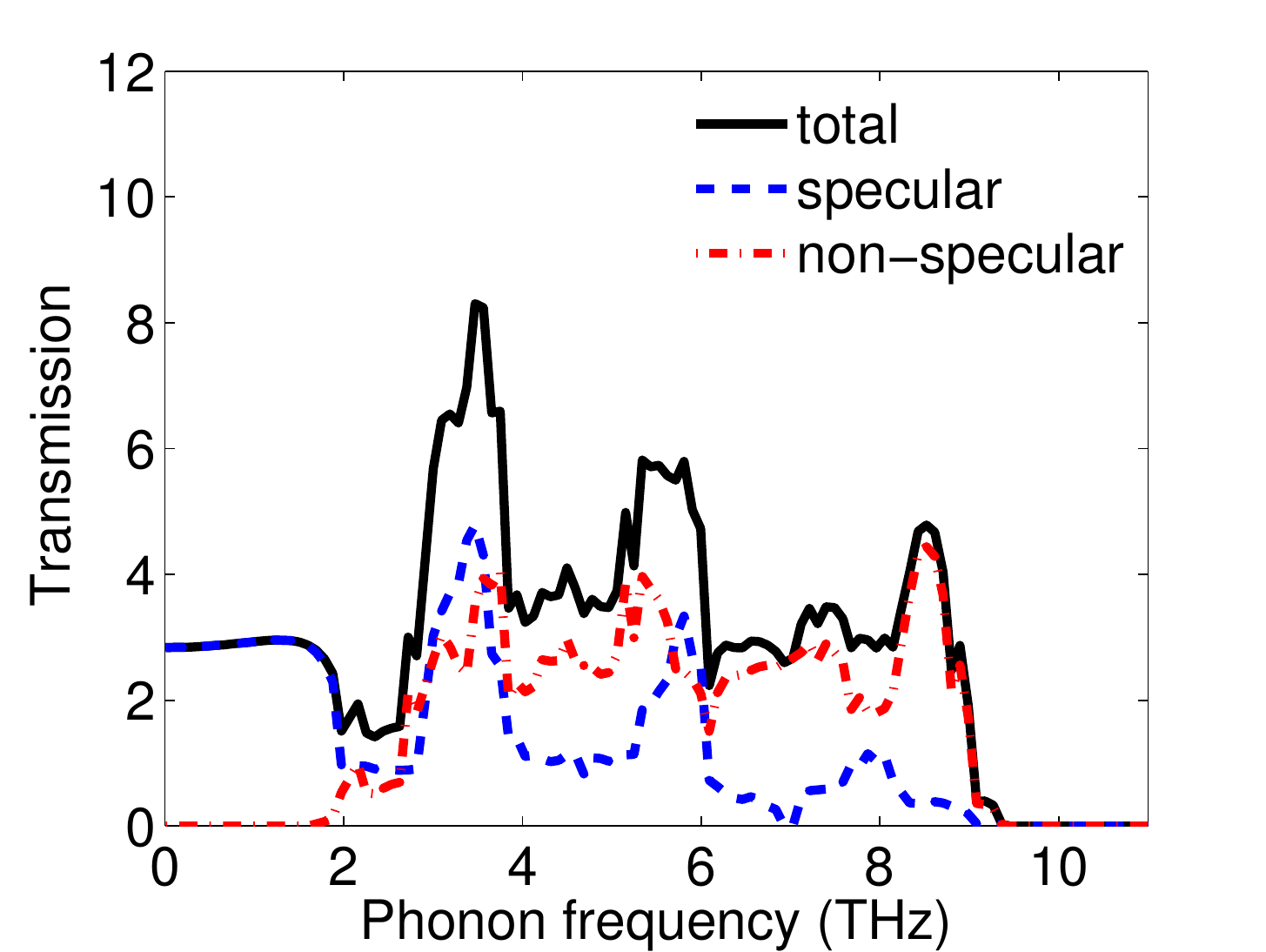}}
\subfloat[]{\includegraphics[height=70mm]{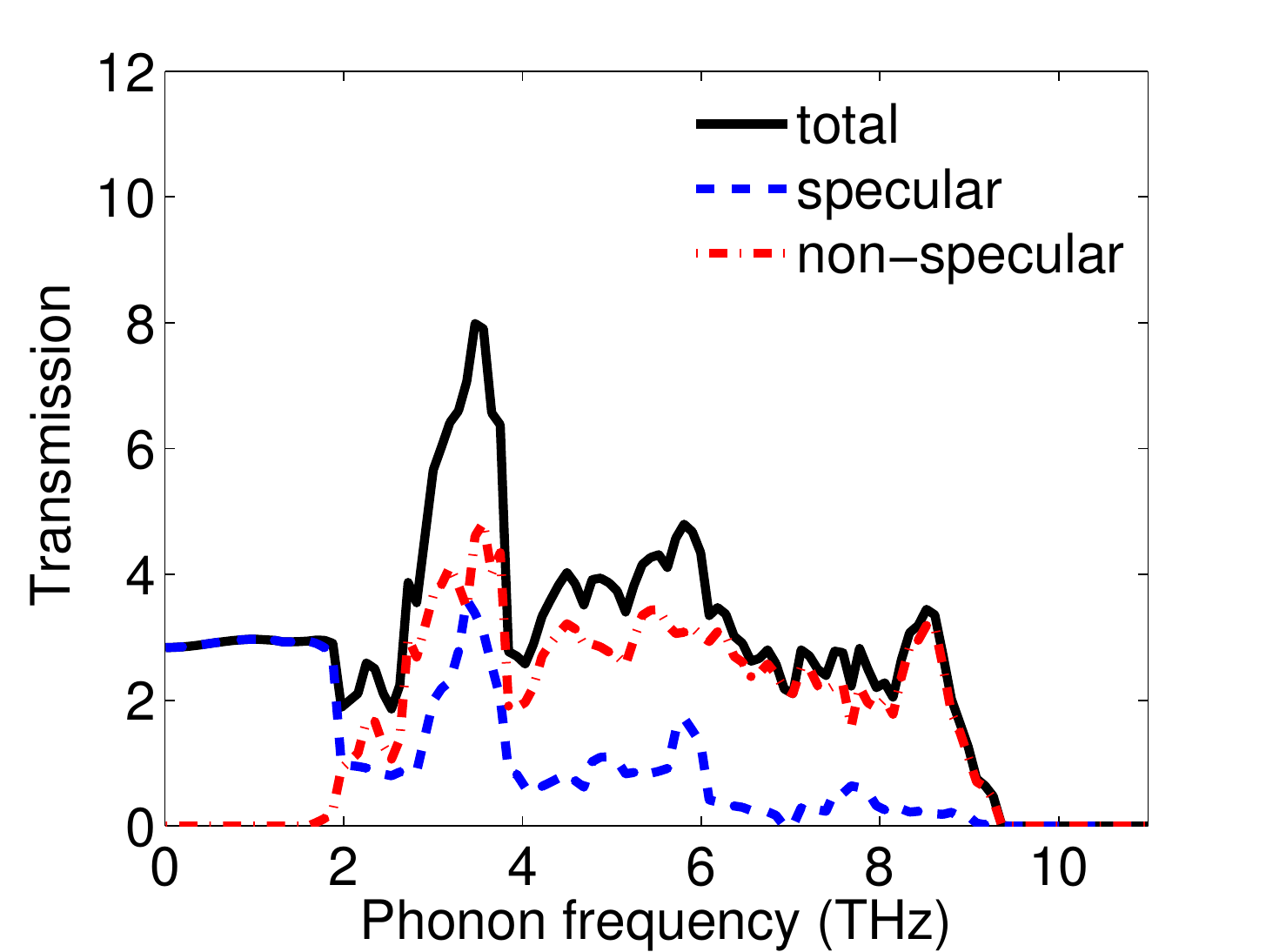}}
\caption{Transmission function across Si-Ge interface for the transverse wavevector at $\Gamma$ point of the supercell BZ. The black curve gives the total transmission function, the blue curve gives the sum of diagonal elements of the transmission matrix in Eq.~(\ref{trans_matrix}) and the red curve corresponds to the sum of off-diagonal elements of the transmission matrix in Eq.~(\ref{trans_matrix}). a) Smooth interface. b) Two atomic layers of intermixing between Si and Ge atoms. c) Four atomic layers of intermixing between Si and Ge atoms. d) Six atomic layers of intermixing between Si and Ge atoms.}\label{trans_spec}
\end{figure}

The mode-resolved approach also enables an in-depth analysis of mode-conversion due to atomic intermixing. As an example, Figure \ref{trans_spec} shows the total transmission function from Si to Ge along with the transmission function between modes of the same in-plane wavevector (termed as `specular' in Figure \ref{trans_spec}) and the transmission function between modes of different in-plane wavevectors (termed as `non-specular' in Figure \ref{trans_spec}). The dissection of the total transmission function into specular and non-specular parts could be understood with a transmission matrix as follows:
\begin{equation}
\begin{blockarray}{cccccccc}
{} & 0 & \vec{b}_1 & (\vec{b}_1+\vec{b}_2) & (\vec{b}_1+2\vec{b}_2) & \dots & \dots & (3\vec{b}_1+3\vec{b}_2)\\
\begin{block}{c(ccccccc)}
  0 & \circ & \circ & \circ & \circ & \circ & \circ & \circ \\
  \vec{b}_1 & \circ & \circ & \circ & \circ & \circ & \circ & \circ \\
  \vec{b}_1+\vec{b}_2 & \circ & \circ & \circ & \circ & \circ & \circ & \circ \\
  \vec{b}_1+2\vec{b}_2 & \circ & \circ & \circ & \circ & \circ & \circ & \circ \\
  \vdots & \circ & \circ & \circ & \circ & \circ & \circ & \circ \\
  \vdots & \circ & \circ & \circ & \circ & \circ & \circ & \circ \\
  3\vec{b}_1+3\vec{b}_2 & \circ & \circ & \circ & \circ & \circ & \circ & \circ \\
\end{block}
\end{blockarray}
\label{trans_matrix}
\end{equation}
$\vec{b}_1$, $\vec{b}_2$ denote the reciprocal lattice vectors of the supercell BZ, and the rows and columns of the above matrix correspond to modes of Si and Ge at the $\Gamma$ point of the supercell BZ. Since the $\Gamma$ point of the $4\times4$ supercell BZ includes 16 different q-points of the extended BZ, the above matrix gives the transmission probability for all these different permutations of mode conversion. The transmission function obtained from conventional AGF provides only the sum of all elements in the above matrix. The diagonal elements of the transmission matrix produce no change in the transverse wavevector while the off-diagonal elements correspond to non-specular transmission with modification of the transverse wavevector. 

Figure \ref{trans_spec}a confirms the intuition that phonon transmission is entirely specular for a smooth interface where the total transmission function is given by the sum of diagonal elements in the transmission matrix. However with intermixing between Si and Ge atoms at the interface (see Figures \ref{trans_spec}b,c,d), the contribution from off-diagonal elements of the transmission matrix, i.e., non-specular interface scattering, increases. The low-frequency or long-wavelength phonons transmit in a specular fashion, since no modes exist for non-zero wavevectors at this frequency. However, the higher frequency modes ($\omega=6-9$ THz) are transmitted almost completely in a non-specular manner for the structure with six layer atomic intermixing (Figure \ref{trans_spec}d). Also, the total transmission function at these high frequencies decreases with increasing levels of intermixing. 

In the mid-frequency range, the transmission function is enhanced at certain frequencies ($\omega=3-5$ THz) for the intermixed atomic structure in comparison to the ideal interface. This enhancement in the transmission function can be understood from a finer mode-resolved analysis of the transmission function corresponding to one of the rows of the transmission matrix in Eq.~(\ref{trans_matrix}). Each row or column of the transmission matrix can be further decomposed into acoustic/optical modes, and the present formulation enables the computation of transmission functions between each pair of such modes. 

\begin{figure}
\centering
\subfloat[]{\includegraphics[height=70mm]{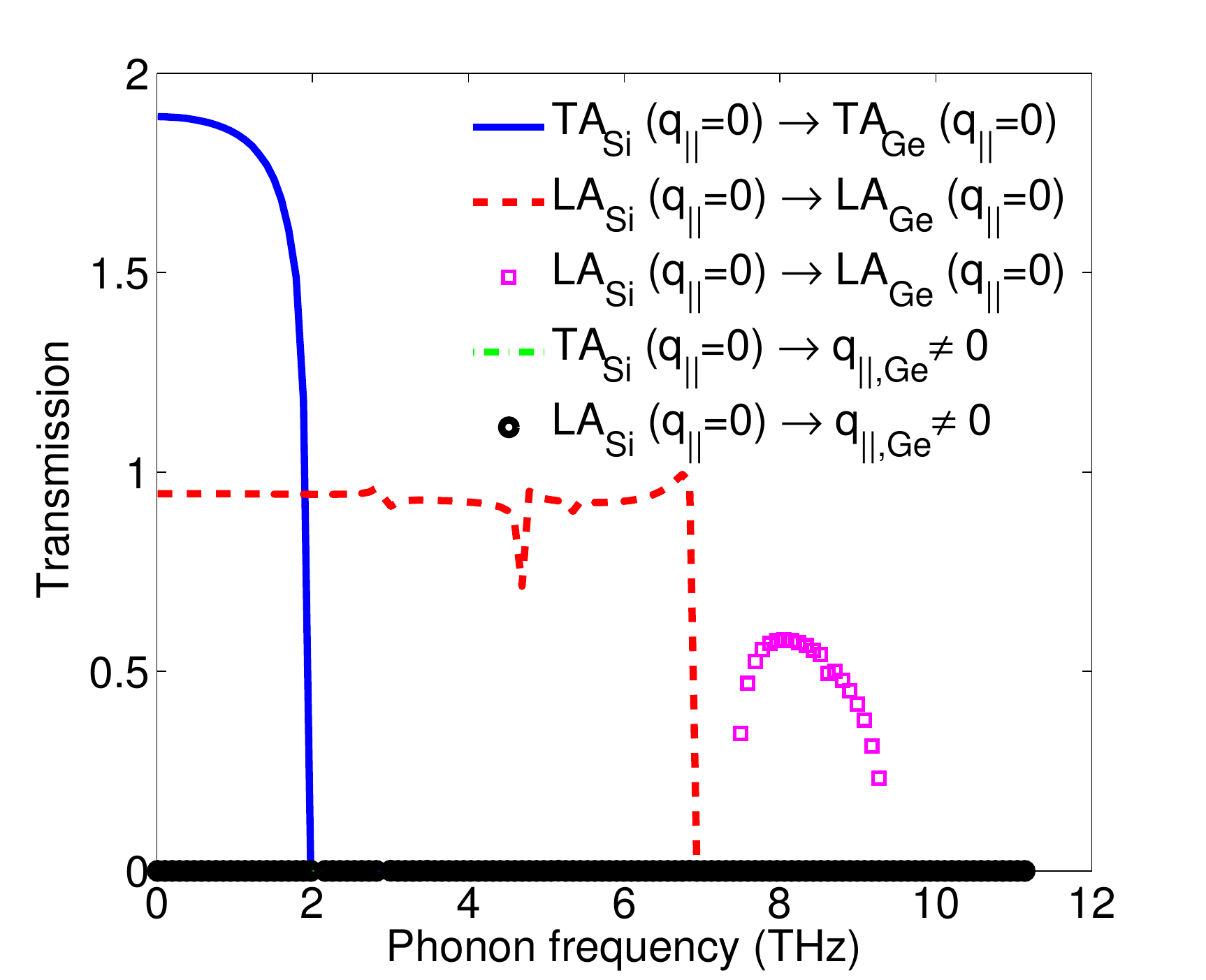}}
\subfloat[]{\includegraphics[height=70mm]{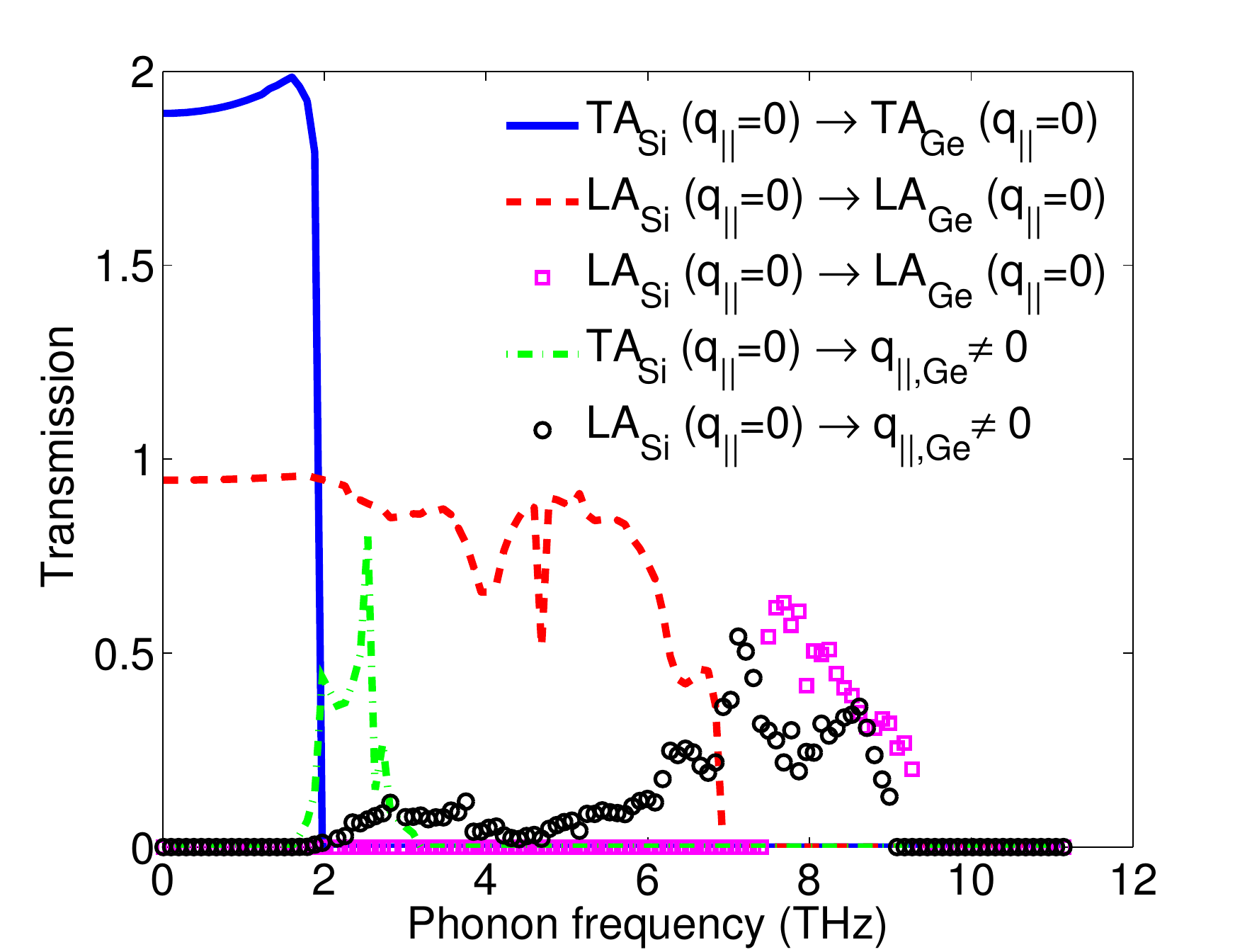}}\\
\subfloat[]{\includegraphics[height=70mm]{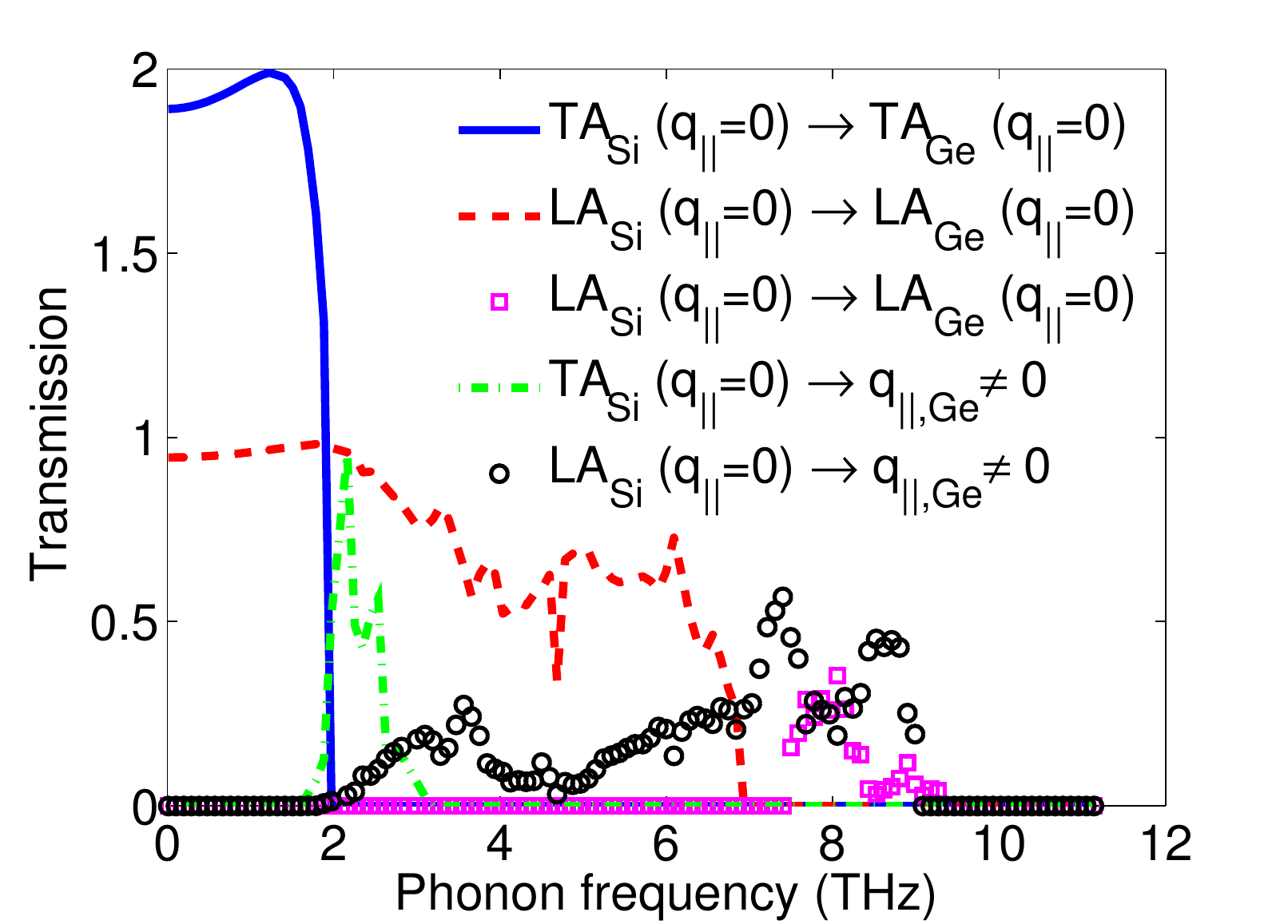}}
\subfloat[]{\includegraphics[height=70mm]{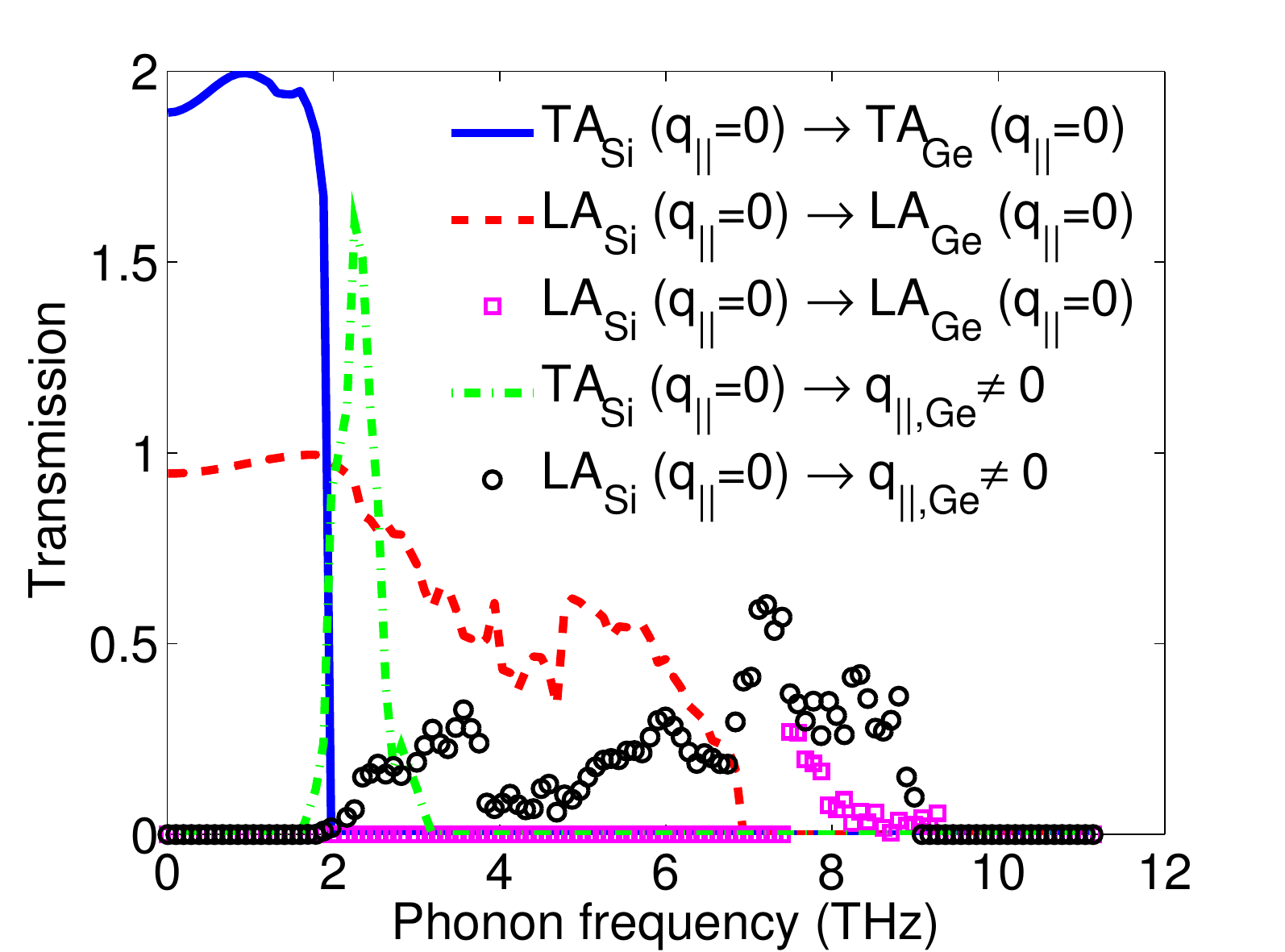}}
\caption{Transmission probabilities for transmission of normally incident TA and LA modes of Si into normally transmitted acoustic and optical modes of Ge along with non-specular transmission probabilities (transmitting into non-zero transverse wavevectors of Ge). a) Smooth interface. b) Two atomic layers of intermixing between Si and Ge atoms. c) Four atomic layers of intermixing between Si and Ge atoms. d) Six atomic layers of intermixing between Si and Ge atoms.}\label{mode_conv}
\end{figure}

As an example, Figure \ref{mode_conv} shows the phonon transmission function for transverse acoustic (TA) and longitudinal acoustic (LA) Si modes incident normally ($\boldsymbol{q}_{t}=0$) on the interface. At low frequencies, we observe a near-perfect transmission of normally incident TA and LA modes of Si into normally transmitted TA and LA modes of Ge. For an ideal interface, the normally incident TA modes of Si have zero transmission beyond the cutoff frequency of TA modes of Ge with $\boldsymbol{q}_{t}=0$ ($\omega \sim 2$ THz). However for a two layer intermixed interface, the higher frequency TA modes of Si with $\boldsymbol{q}_{t}=0$ show significant transmission into modes of Ge with non-zero transverse wavevectors (see green curve in Figure \ref{mode_conv}b). Also, this transmission increases with greater interfacial atomic mixing. Hence, we observe that atomic mixing at the interface increases the number of available modes of Ge to which Si modes can elastically transfer their energy. This increase in the number of allowed states in Ge is a direct consequence of relaxation in the condition for conservation of transverse momentum. 

Similar to TA modes, intermixing also enhances the transmission of LA modes in certain frequency ranges. For example, normally incident LA modes of Si have zero transmission in the bandgap between LA and LO modes of Ge ($\omega = 6.7-7.5$ THz) for a perfect interface with no intermixing (see Figure \ref{mode_conv}a). However, such a bandgap exists only for $\boldsymbol{q}_{t}=0$ (normal transmission), and modes belonging to this frequency range exist in Ge for other in-plane wavevectors which correspond to oblique transmission. Intermixing of atoms at the interface allows the LA modes of Si to transmit into these oblique modes (see black circles in Figure \ref{mode_conv}b,c,d). 

\begin{figure}
\centering
\subfloat[]{\includegraphics[height=70mm]{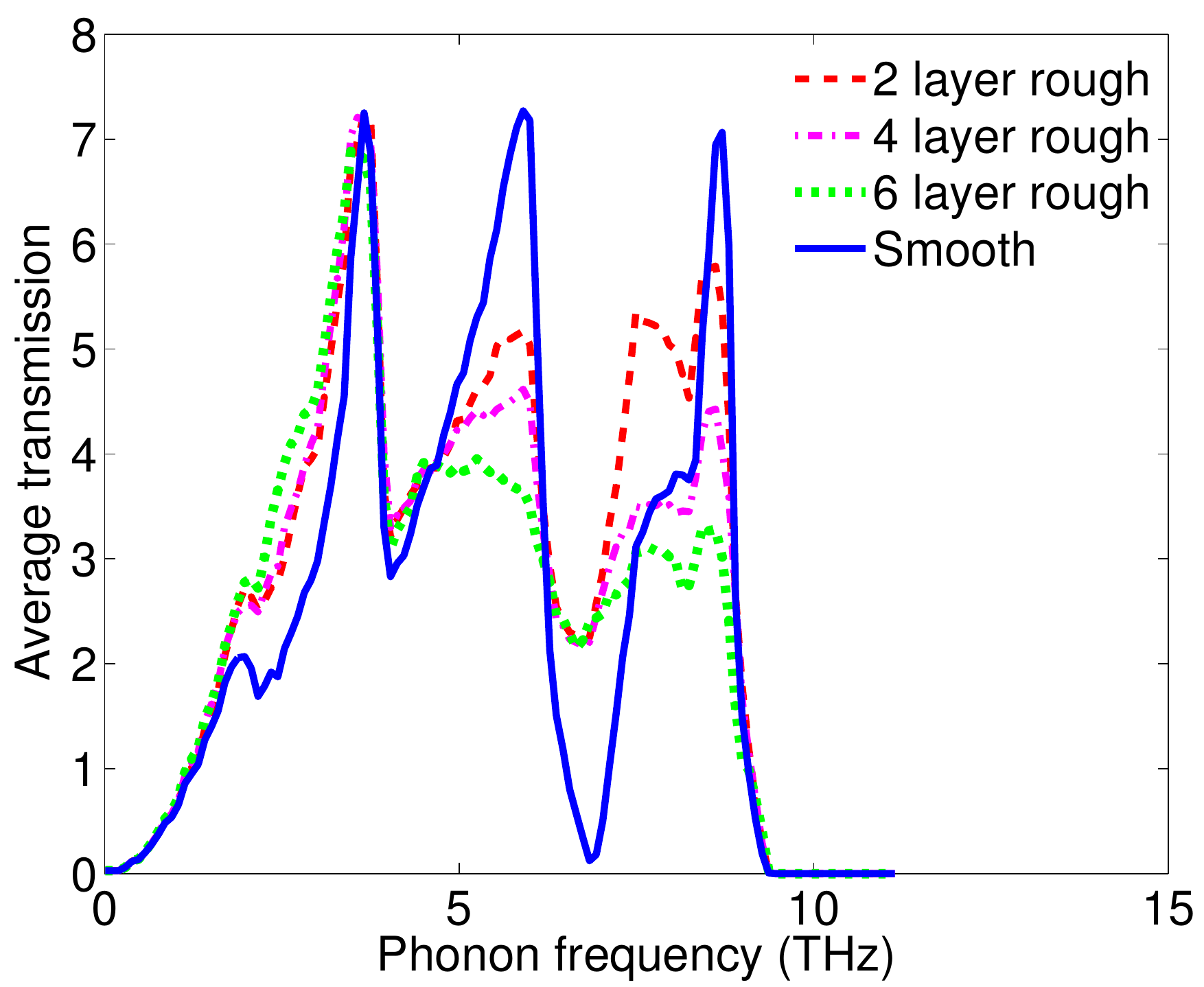}}
\subfloat[]{\includegraphics[height=70mm]{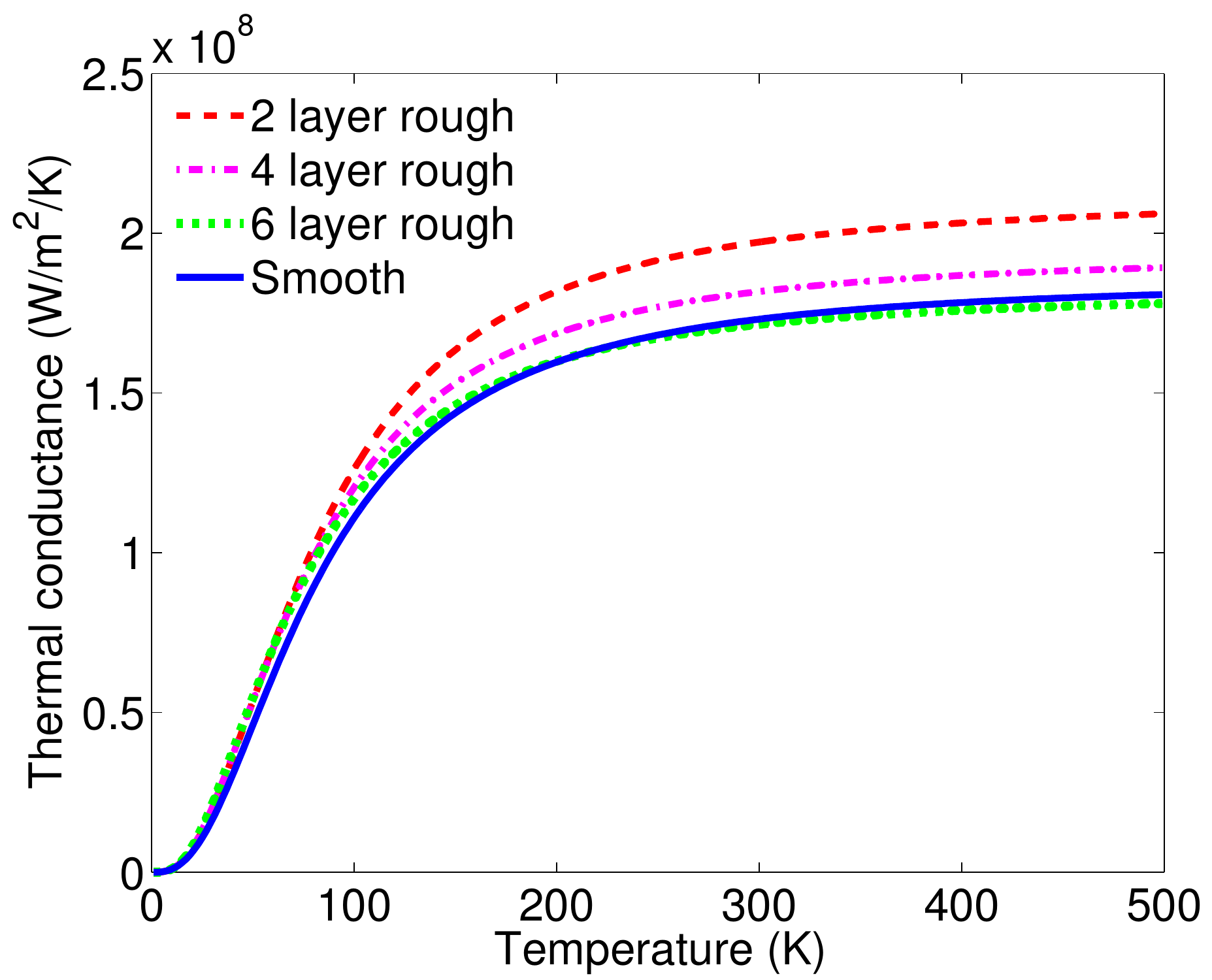}}
\caption{a) Average transmission function obtained by averaging over 100 transverse wavevectors in the BZ of the $4\times 4$ in-plane supercell. b) Total thermal interface conductance of Si-Ge interface for varying levels of interfacial intermixing of atoms.}\label{tot_cond}
\end{figure}

Although intermixing of atoms increases the number of permitted mode conversions across the interface, interfacial disorder also reduces the overall transmission function at certain frequencies. Figure \ref{tot_cond}a shows the transmission function averaged over 100 transverse wavevectors in the BZ of the $4\times 4$ supercell. We observe that the overall transmission function increases with greater intermixing of atoms for low frequencies ($\omega = 1.5-3$ THz) while the transmission function decreases with intermixing for phonons in the frequency range 4.5-6 THz. Figure \ref{tot_cond}a also reveals that phonons in the frequency range of 6-8 THz have higher transmission (compared to the perfect interface) for the structure with two layers of intermixing while the transmission function decreases with further mixing of atoms. A similar trend is observed in the total thermal interface conductance which is maximized for two atomic layers of intermixing while the six layer intermixed interface has nearly the same conductance as the pristine interface. 

Results in Figure \ref{tot_cond}b point to the existence of an optimum level of interfacial mixing to maximize thermal interface conductance. The existence of such an optimum mixing length can be attributed to competing effects of enhancement in transmission function due to higher number of phonon modes that are allowed to exchange energy and a decrease in the total transmission function due to interfacial disorder. Similar results were reported in ref.~\cite{tian2012enhancing} where the enhancement in transmission was attributed to a smooth transition in phonon DOS across an intermixed interface as compared to an abrupt transition of phonon DOS in a smooth interface. The mode-resolved AGF approach developed in this paper however elucidates the mechanics of mode conversion and the degrees of freedom available for interfacial scattering of each incident phonon mode.

\section{Conclusions}
We have presented a reformulation of the surface Green's functions within the conventional atomistic Green's function method that enables the computation of mode-resolved phonon transmission functions. The Sancho-Rubio or decimation method is the commonly used technique for calculation of surface Green's functions and contact self-energies. However, the Sancho-Rubio method does not allow the calculation of mode-resolved surface Green's functions that are needed to determine mode-resolved phonon transmission functions. The alternative method developed in this paper is based on the use of Dyson and Lippmann-Schwinger equations to obtain surface Green's functions directly from bulk phonon eigenvectors. This approach enables a straightforward calculation of each phonon mode's contribution to the surface Green's function and contact self-energy. The only modification required for the calculation of mode-resolved transmission functions is the use of an alternative mode-resolved technique to compute surface Green's functions. The rest of the algorithm remains the same as the conventional AGF method with the use of the Caroli formula to compute transmission functions; the self-energies used in the Caroli formula are however mode-resolved self-energies computed with the technique presented in this paper. 

We demonstrated the proposed technique through analysis of thermal transport on Si-Ge interfaces with varying levels of intermixing between atoms. Our results reveal that interfacial intermixing relaxes the condition on conservation of transverse momentum and allows for increased degrees of freedom for elastic transfer of energy between bulk Si and bulk Ge phonon modes. The increased phase space for elastic scattering results in a higher transmission function in some phonon frequency ranges and leads to an increase in interfacial thermal conductance for intermixed interfaces in comparison to ideal or smooth interfaces. The example studied in this paper demonstrates the usefulness of the proposed extension to conventional AGF by providing new insights into the microscopic mechanisms of interfacial phonon scattering. More broadly, the present approach can provide mode-resolved transport as inputs to multiscale models such as the Boltzmann transport equation for studying heat transport at mesoscopic length scales. 
\section*{Acknowledgements}
SS acknowledges financial support from the Office of Naval Research (Award No: N000141211006) and Drs. Helen and Marvin Adelberg fellowship from the School of Mechanical Engineering at Purdue University. UVW acknowledges funding from an AOARD grant FA2386-15-1-0002 and support from a JC Bose National Fellowship and Nano Mission of the Department of Science and Technology.

\bibliographystyle{ieeetr}
\bibliography{references}
\end{document}